\documentclass[slac_one]{revtex4}
\usepackage{graphicx}
\usepackage{fancyhdr}
\begin{document}

\title{SUSY Searches in Multi-lepton and Di-photon Final States at D\O } 

%

\author{James Kraus (for the D\O\ Collaboration)}
\affiliation{Michigan State University, East Lansing, MI 48224, USA}
%

\begin{abstract}
We report results from searches for supersymmetry with trileptons 
plus missing transverse energy and diphotons plus missing transverse energy
made using the D\O\ detector at the Fermilab Tevatron
collider.  These searches are optimized for MSSM SUSY and GMSB, respectively.
New limits are set on the considered SUSY models.

\end{abstract}

\maketitle

\thispagestyle{fancy}


\section{INTRODUCTION AND SUSY THEORY} 
The standard model of particle physics has proved remarkable successful in 
explaining the behavior of particles thus far observed in high energy physics 
experiments.  However, the standard model requires significant fine tuning to 
explain why the scale of electroweak symmetry breaking is so much less than 
the GUT scale.  Supersymmetry (SUSY) theory provides a natural explanation for 
this by proposing a supersymmetric parter for each standard model particle.  In the
Minimal Supersymmetric extensions of the Standard Model (MSSM), there are hundreds of 
free parameters, so the searches described below set limits on specific models with 
some parameters held fixed.  These searches are for charginos, $\tilde{\chi}^+$, which are linear 
combinations of the SUSY partners of the $W$ and charged Higgs, and neutralinos, $\tilde{\chi}^0$,which are 
linear combinations of the SUSY partners of the $Z$, $\gamma$, and neutral Higgs bosons.
R-parity conservation is assumed, so that the lightest SUSY particle (LSP) is stable.~\cite{snowmasspaper1,snowmasspaper2}
In minimal supergravity (mSUGRA) models, the only free parameters are ``the scalar mass parameter $m_0$,
the gaugino mass parameter $m_{1/2}$, the trilinear coupling $A_0$, the ratio of the Higgs 
vacuum expectations values, $\mbox{tan} \beta$, and the sign ofthe supersymmetric Higgs 
mass parameter, $\mu$.''~\cite{snowmasspaper2}
In gauge-mediated symmetry breaking (GMSB) models, chiral messenger particles mediate the symmetry breaking, leading 
to the free parameters in the theory 
being $N_m$, the number of messengers, $M_m$, the masses of those messengers, $\Lambda$, the apparent scale of the 
symmetry breaking at low energy, $C_{grav}$ which sets the mass for the gravitino (the LSP in GMSB), the sign of 
$\mu$, and $\mbox{tan} \beta$.  Minimal models set $C_{grav}=1$~\cite{snowmasspaper1, snowmasspaper2, gmsbfermipaper}.

\section{DETECTOR}
``The D\O\ detector \cite{nim1, nim2, nimmu} contains tracking, calorimeter and muon
subdetector systems.  Silicon microstrip tracking detectors (SMT) near the 
interaction point cover pseudorapidity $|\eta| < 3$ to provide tracking and vertexing 
information.  The central fiber tracker (CFT) surrounds the SMT, providing 
coverage to about ($|\eta|=2$).  The CFT has eight concentric cylindrical layers of overlapped scintillating 
fibers providing axial and stereo ($\pm 3^\circ$) measurements.  A 2T solenoid surrounds these tracking 
detectors.  Three uranium-liquid argon calorimeters measure particle energies.  The central calorimeter (CC) 
covers $|\eta| < 1$, and two end calorimeters (EC) extend coverage to about $|\eta|=4$.  The calorimeter is 
highly segmented along the particle direction, with four electromagnetic (EM) and for to five hadronic 
sections, and transvers to the particle direction with typically $\Delta \eta = \Delta \phi = 0.1$, where $\phi$ 
is the azimuthal angle.  The calorimeters are supplemented with central and forward scintillating strip
preshower detectors (CPS and FPS) located in front of the CC and EC.  Muons are measured just 
outside the calorimeters, and twice more outside the 1.8T iron toroidal magnets, over the 
range $|\eta|<2$.  
Scintillators surrounding the exiting beams allow determination of the luminosity.
A three level trigger system selects events for data logging at about 100 Hz.''~\cite{Abazov:2008yf}

\section{DIPHOTON SEARCH}

The di-$\gamma$ plus missing transverse energy (MET) search is optimized for GMSB~\cite{GMSBref}.  It is assumed that R-parity is conserved and that the next lightest supersymmetric particle (NLSP) is the neutral chargino.  The NLSP is pair produced, and decays to a photon and a stable gravitino that escapes the detector undetected, showing up as missing energy.  The data used in this search was taken during Run IIa of D\O\ and has an integrated luminosity of 1.1 fb$^{-1}$.
 
There are three potential sources of background in this search.  First, there are events with real di-photons and MET, which primarily consists of $Z \gamma \gamma \rightarrow \nu \nu \gamma \gamma$ and $W \gamma \gamma \rightarrow l \nu \gamma \gamma$, where the lepton is lost or mis-identified.  The next source is events with real MET and a one or two fake photons.  These events are mostly $W \gamma \rightarrow e \nu \gamma$ and $W j \ \rightarrow e \nu jet$, where the electron and or jet fake the photon.  The last source of background is events with fake MET, which consists of QCD di-photon and di-jet events where the energy is mismeasured and the jets fake photons.

We look for photons only in the CC.  They must have a narrow energy deposition with 96\% of their energy within the EM calorimeter.  The probability of a track from the CFT matching the photon must be less than 0.001, and only $\gamma$ with an $E_T > 25$ GeV are considered.  Also, the photon must be isolated in both the tracker and the calorimeter.

The anti-track match cut is enhanced by looking for hits in the CFT where a track would have been expected to pass if the EM cluster were due to an electron.   Doing so effectively improves the electron tracking efficiency from $93.0\% \pm 0.1\%$ to $98.6\% \pm 0.1\%$.

One major innovation used in this analysis is that hits in the CPS detector are matched to the photon EM cluster and used to extrapolate the z position of the photon at the beamspot.  This is important because the CPS has a much finer resolution than the EM calorimeter, so the pointing is much more precise. The $z$ resolution at the beamline for a CPS matched photon has been found to be 2.3 cm from $Z \rightarrow e e \gamma$ data.  This analysis requires that at least one of the 2 photons have a CPS hit, and imposes a vertex cut between the primary vertex of the event and the photon of $\Delta z < 10$ cm for events with one CPS matched photon and $\Delta z < 7$ cm with two CPS matched photons.  This improves the MET resolution, and reduces the number of events where the 2 photons come from different vertices.

\begin{figure}[!ht]
\includegraphics[width=67mm]{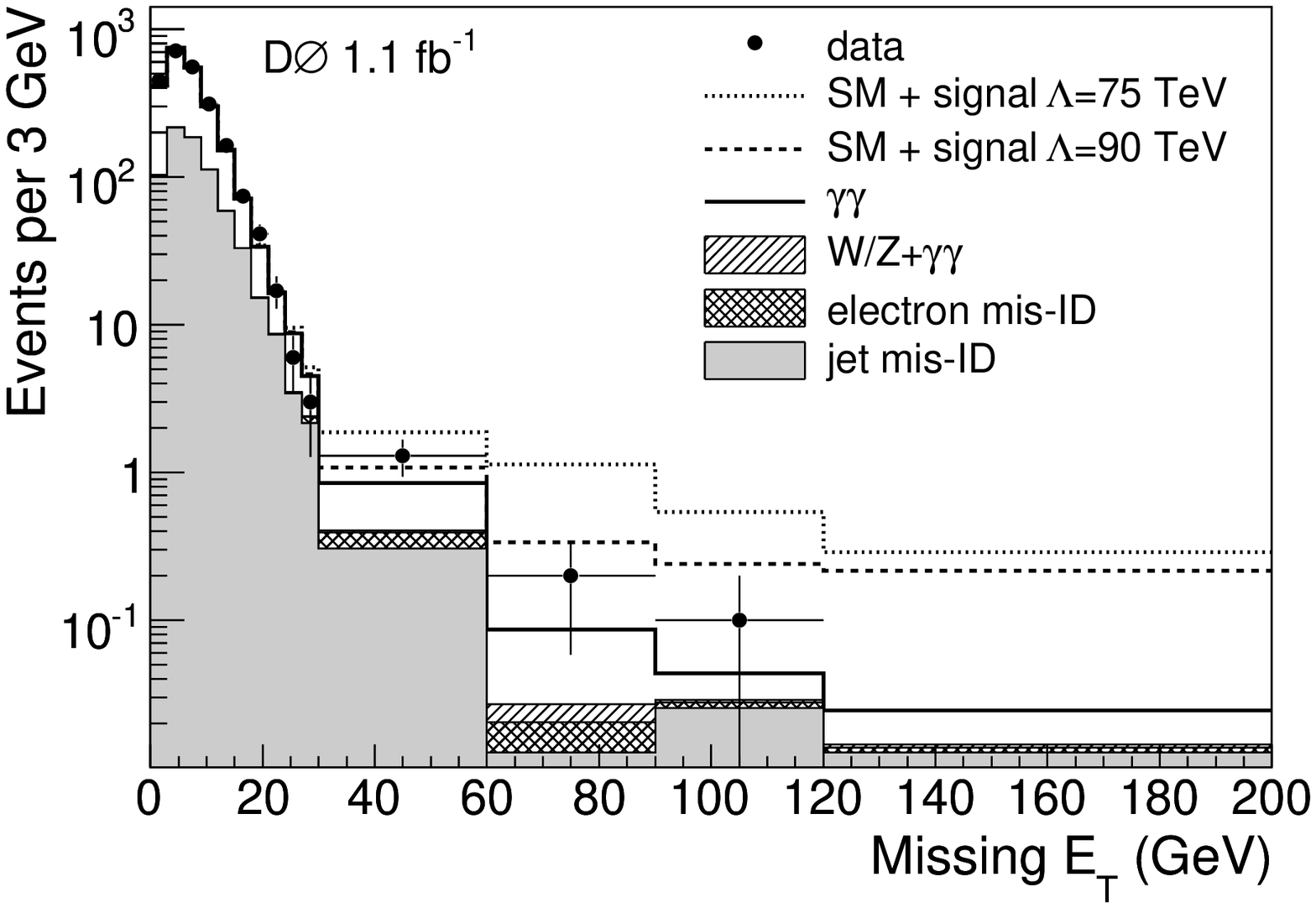}%
\includegraphics[width=67mm]{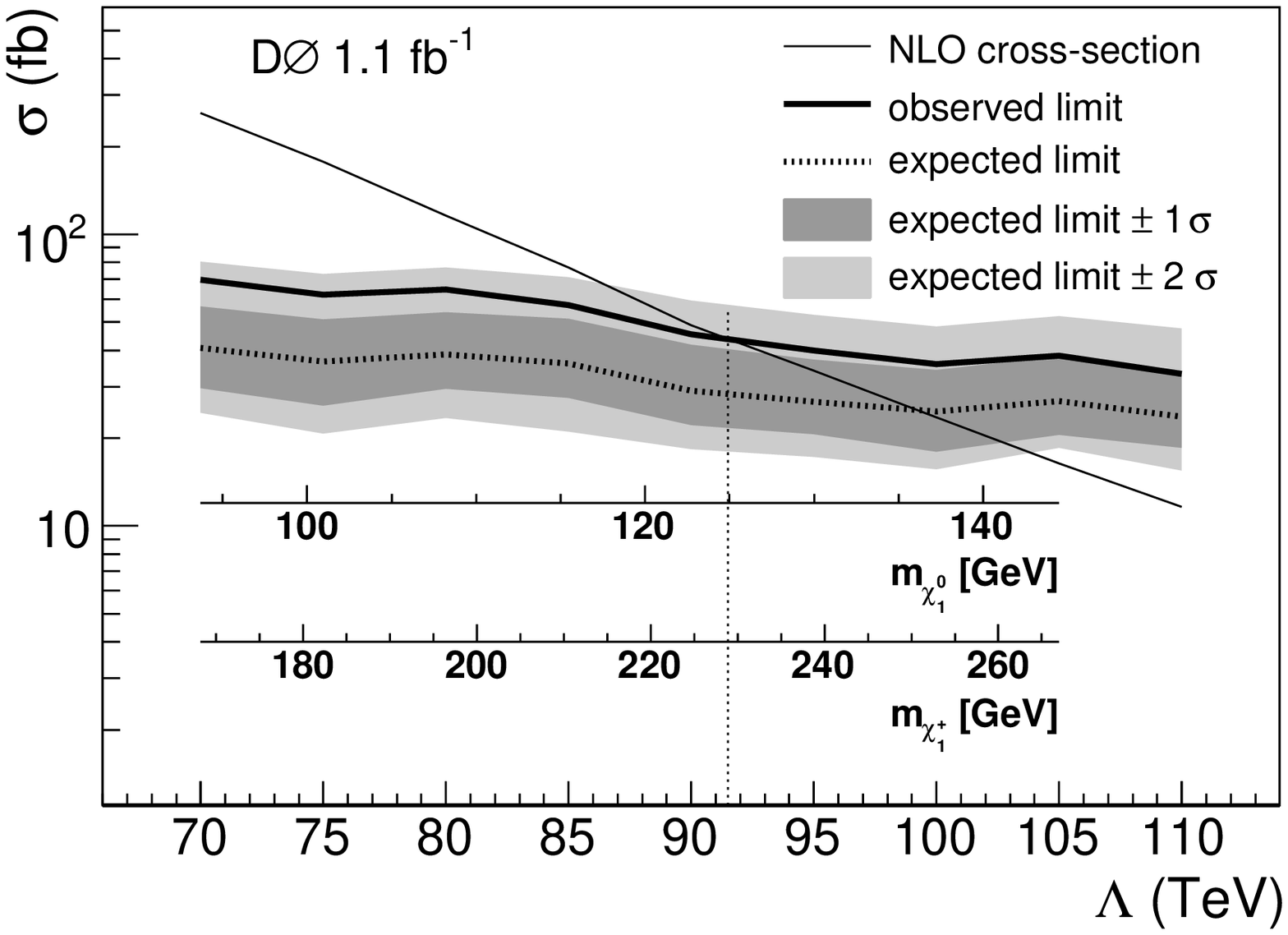}%
\caption{ Left: Missing $E_T$ distribution for diphotons.  Right: Diphoton limit set on the value of $\Lambda$ versus di-photon cross section.  Chargino and Neutralino mass limits also shown. \label{GMSBMET}}
\end{figure}
The $W \gamma \gamma$ and $Z \gamma \gamma$ backgrounds are estimated from COMPHEP MC and found to be $0.10 \pm 0.04$ and $0.15 \pm 0.06$ events, respectively.  

The $W \gamma \rightarrow e \nu \gamma$ and $W j \rightarrow e \nu jet$ fake rates are measured by finding $W \gamma$ events in data and multiplying by $\frac{1-\epsilon}{\epsilon}$, where here $\epsilon$ is the efficiency for reconstructing and matching the electron's track to the electron's calorimeter stub.  

The backgrounds with fake MET are determined using two different methods.  The first method is to reverse some of the quality cuts on the photons, resulting in a sample of jets that are very similar to photons.  The other method is to look at MET in $Z \rightarrow ee$ events, as no real MET is expected in these events and the energy deposition of the electrons is similar to that of photons.  The MET distribution produced via the two methods agrees within errors.  The fake MET background is normalized to the signal by scaling the MET distributions at $\mbox{MET} < 12$ GeV to match that of the good di-$\gamma$ distribution below 12 GeV.

Comparison of the MET distribution from data and expected backgrounds is shown in figure ~\ref{GMSBMET}.  There is no indication of an excess at large MET that would indicate a GMSB signal.

For the purpose of setting limits, we generate MC based on GMSB Snowmass Slope 8 ~\cite{snowmasspaper1, snowmasspaper2}, with $N_m = 1$, $\mu > 0$, $\mbox{tan} \beta = 15$, and $M_m / \Lambda = 2$.  In this scenario, we set a limit on the scale of SUSY symmetry breaking, $\Lambda > 91.5 \mbox{ TeV}$ at the 95\% CL.  This corresponds to a limit on the chargino mass of 229 GeV and neutralino mass of 125 GeV in this scenario at the 95\% CL.  The limit is plotted against the Snowmass slope in figure~\ref{GMSBMET}.

\section{TRILEPTON SEARCH}
\begin{figure}[ht]
\includegraphics[width=67mm]{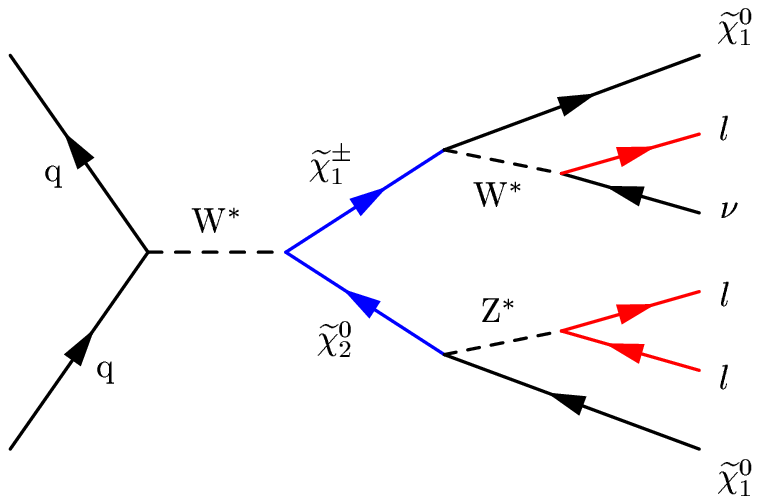}%
\includegraphics[width=67mm]{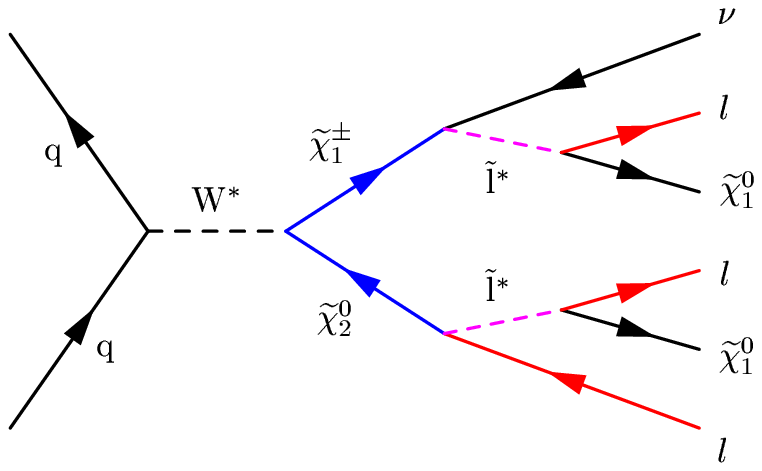}%
\caption{ Leading order Feynman diagrams for weak chargino/neutralino production and decay to trileptons plus MET \label{feyndiag}}
\end{figure}

The trilepton SUSY search is optimized to look for MSSM.  R-parity conservation is assumed, and the neutralino $\tilde{\chi}^{0}_{1}$ is assumed to be the LSP.  This search looks for the decays of weakly produced $\tilde{\chi}^{\pm}$ $\tilde{\chi}^{0}$ pairs to three leptons plus MET, as shown in figure~\ref{feyndiag}.  Three trilepton search channels are described here:  $eel$ ~\cite{eelpaper}, $\mu \mu l$ ~\cite{mumulpaper}, and $ e \mu l$ ~\cite{mumulpaper}.  A fourth search not detailed here, $\mu^{+} \mu^{+}$~\cite{mumupaper}, is designed to 
fills in the gap in phase space where the mass difference between the lightest slepton and chargino masses is so small that the third lepton is too soft to be found at  D\O.  

The $\mu \mu l$ and $e \mu l$ analyses use 1 fb$^{-1}$ of integrated luminosity from Run IIa of D\O.  The $e e l$ analysis uses 1.7 fb$^{-1}$ of integrated luminosity from Run IIa and RunIIb of D\O.

Electron ID for the trilepton search is similar to that of $\gamma$ ID in the GMSB search detailed above, except that a track match is required.  The muon ID requires a track in the CFT that is isolated in both the tracker and calorimeter and matched to a hit in the muon chambers.  In each analysis, the third lepton is a track that is isolated in both the tracker and the calorimeter.  This track must have at least 17 SMT+CFT hits or at least 14 CFT hits.

Backgrounds for this analysis come from vector boson plus fake lepton events, diboson events, top pair production, Upsilon to dilepton decays, and QCD multijet events with the jets faking leptons.  $Z/\gamma^{*}+j \rightarrow llj$ and  $W+j \rightarrow l \nu j$ are modeled using Pythia+Alpgen MC.  The $WW$, $WZ$, $W/\gamma$, $ZZ$, $t \overline{t}$ and $\Upsilon \rightarrow ll$ backgrounds are modeled with Pythia.  The QCD multijet backgrounds are modeled by reversing some of the lepton quality cuts in data.

The SUSY signal is generated with version 6.319 of Pythia.  The optimization is done using the 3 lepton max scenario, where the mass difference between the second neutralino and lightest slepton is small enough to disfavor the decay of the $\tilde{\chi}^{0}_{2}$ to anything other than an $e$ or $\mu$.  We use SUSY parameters $\mbox{tan} \beta = 3$, $A_{0} = 0$, and $\mu > 0$.  SUSY points are generated for chargino masses between 98 and 150 GeV.
\begin{figure}[ht]
\includegraphics[width=67mm]{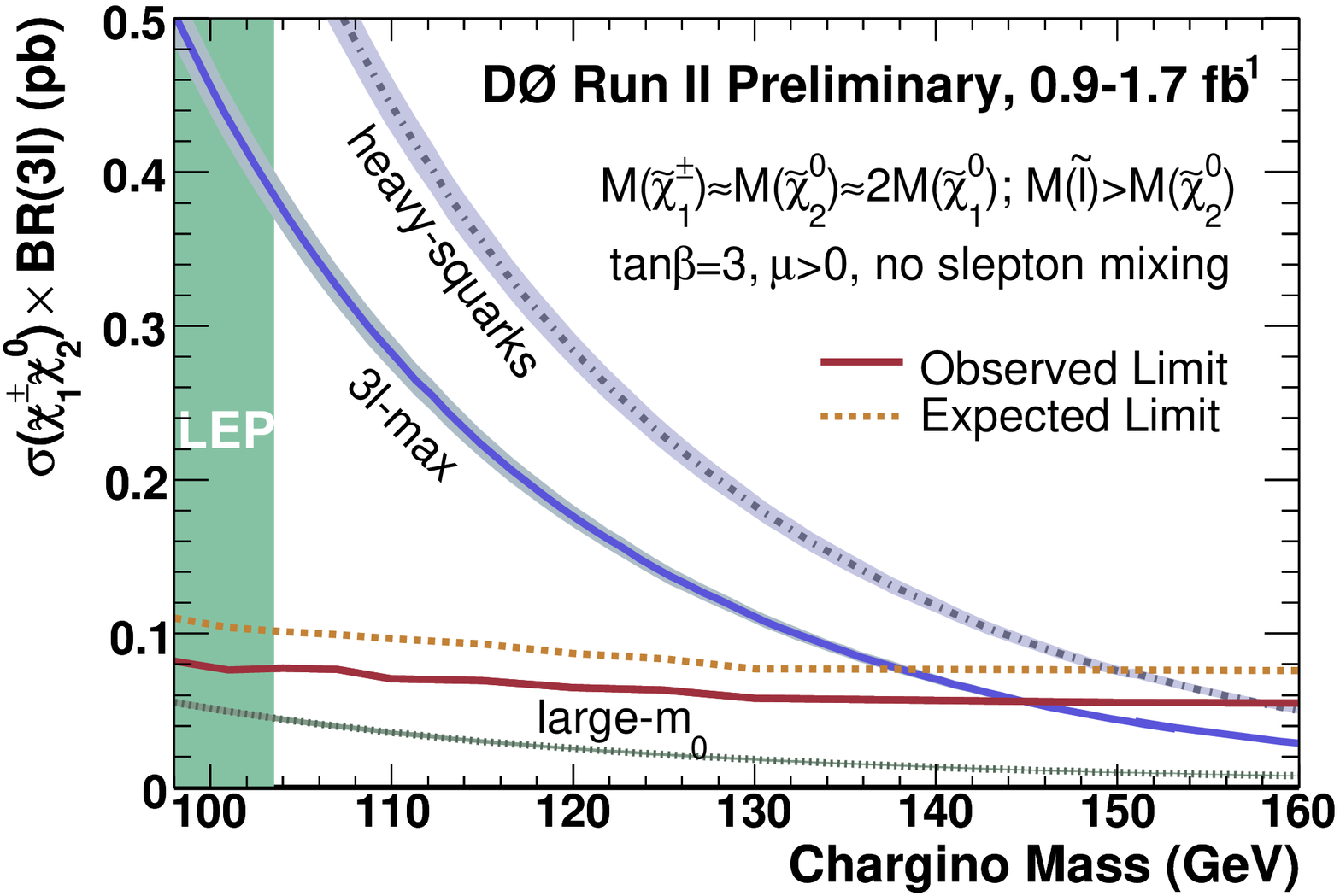}%
\includegraphics[width=67mm]{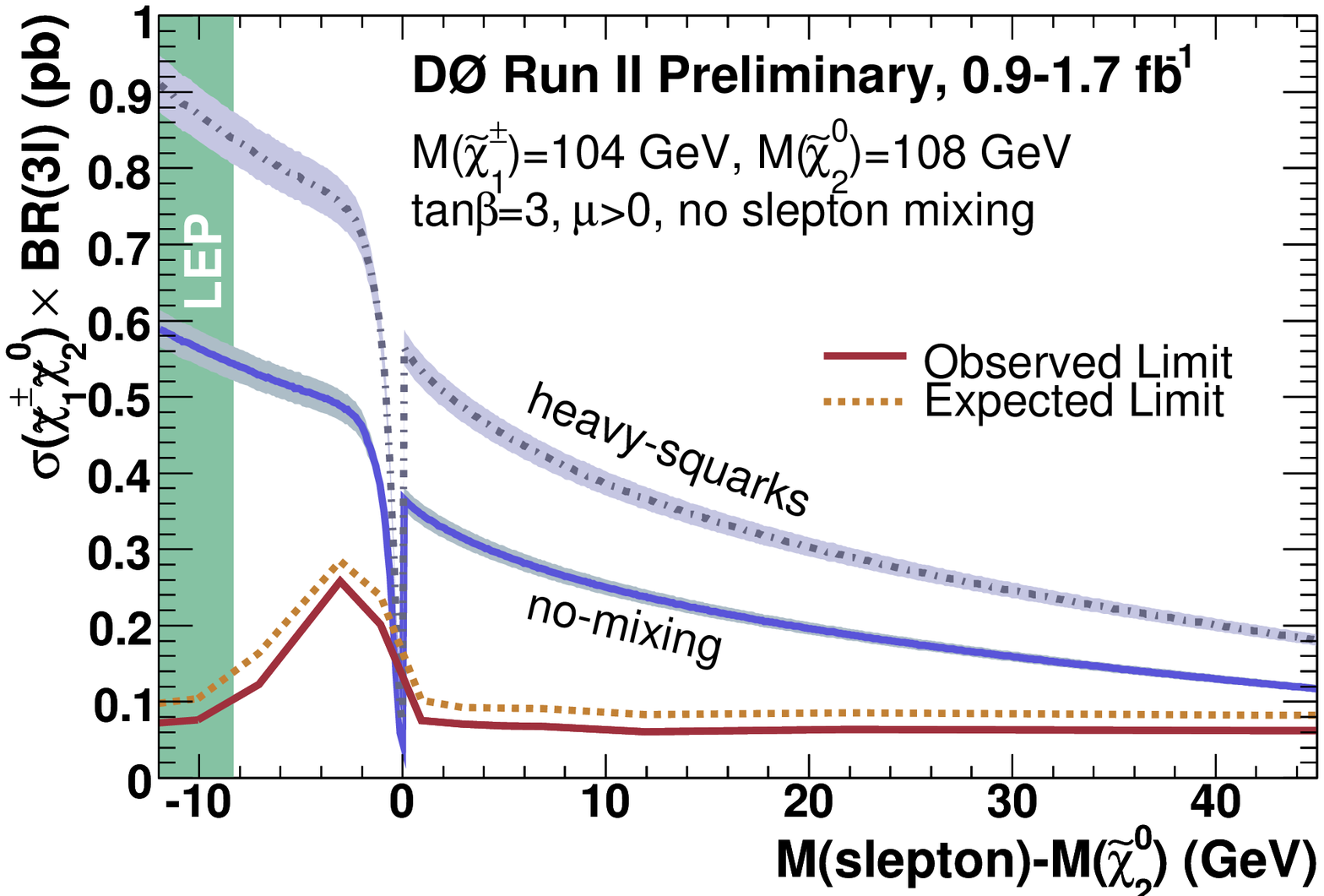}%
\caption{ Trilepton SUSY cross section limits.  Left:  Cross section limit vs. chargino mass where the mass of the lightest slepton is greater than the mass of the next-to-lightest chargino.  Right:  Cross section limit vs. $\Delta M(\tilde{l}-\tilde{\chi}^{0}_{2})$ \label{trilepmainresult}}
\end{figure}

The $eel$ analysis places the following cuts on its events.  The objects must all come from the same vertex and have a $\Delta R = \sqrt{\Delta \phi^{2} + \Delta \eta^{2}} > 0.4$ between them.  There are $p_T$ cuts of 12, 8, and 4 GeV on the leading and second electron and isolated track, respectively.  To remove $Z$ events, the invariant mass of the $ee$ must be between 18 and 60 GeV/$c^2$, the mass of the $el$ must be less than 60 GeV/$c^2$, and $\Delta \phi (e,e) < 2.9$.  The MET must be greater than 22 GeV and the significance of the MET must be more than 8.0.  MET significance is defined as 
$\frac{\mbox{MET}}{\sqrt{\sum_{jet} \sigma^{2}_{E^{jet}_{T} \| \mbox{MET}}}}$, where 
$  \sigma_{E^{jet}_{T} \| \mbox{MET}}$ is the component of the error on the jet energy which is parallel to the MET.  The transverse mass of the lead electron plus MET must be more than 20 GeV, and if $M_T > 65$ GeV, the $l$ track must have a $p_T > $ 7 GeV to cut out $W$ plus track background.  To remove $t \overline{t}$ background, we require $H_T$, the scalar sum of the $E_T$ of all jets in the event, to be less than 80 GeV.  Lastly, a $p_T (l) \times \mbox{MET}> 220 \mbox{ GeV}^2$ to get rid of remaining backgrounds.  No candidate events survive the final cut.

The cuts for the $\mu \mu l$ final state are similar to those of the $e e l$ final state.  There is an additional $p_T$ balance cut of $0.3 < (p_T (\mu_1) + p_T(\mu_2) + \mbox{MET})/p_{T}(track) < 3$, as for this variable $WZ$ is close to 0 and QCD and Z backgrounds are generally more than 3.  The 
$p_T (l) \times \mbox{MET}$ cut is set to 150 GeV$^2$, resulting in 2 events surviving all cuts.

In the $e \mu l$ final state, we require MET $> 10$ GeV with a significance of 8.0.  The lower of the transverse mass of the $e$ or $\mu$ and the MET must be between 20 and 90 to remove both events with poorly measured MET and some $WW$ background.  There is a significant background from $W$ plus track events, so the $p_T$ of the $l$ must be greater than 7 GeV if the transverse mass of the track plus MET is between 60 and 90 GeV.  Finally, To remove $WZ$ background, we place an invariant mass cut of between 5 and 70 GeV on the $e l$ and $\mu l$.  This final cut removes all remaining events so that no candidates were found.

These results are combined with the same-sign dimuon result to create the limit plot shown in figure~\ref{trilepmainresult}.  Under the 3 lepton max scenario described above, we place a limit on the chargino mass of 145 GeV at the 95\% CL.  In figure~\ref{trilepmainresult}, we show the limit on chargino/neutralino production where the mass of the lightest slepton is more than or close to the mass of the second lightest neutralino.





\begin{thebibliography}{99} 

\bibitem{snowmasspaper1}
S.P. Martin, S. Moretti, J.M. Qian, and G.W. Wilson, ``Direct Investigation of Supersymmetry: Subgroup
summary report,'' in {\it Proceedings of the APS/DPF/DPB Summer Study on the Future of Particle 
Physics (Snowmass 2001),} edited by N. Graf, eConf {\bf C010630}, p. 346 (2001).
\bibitem{snowmasspaper2}
B.C. Allanach {\it et al.}, Eur. Phys. J. C {\bf 25}, 113 (2002).
\bibitem{gmsbfermipaper}
H.~Baer, P.~G.~Mercadante, X.~Tata and Y.~l.~Wang, Phys.\ Rev.\ D {\bf 60}, 055001 (1999) [arXiv:hep-ph/9903333].
\bibitem{nim1}
S.~Abachi {\it et al.} [D0 Collaboration], Nucl. Instrum. Methods
Phys. Res A {\bf 338}, 185 (1994).
\bibitem{nim2}
V.M.~Abazov {\it et al.} [D0 Collaboration], Nucl. Instrum. Methods
Phys. Res A {\bf 565}, 463 (2006).
\bibitem{nimmu}
V.M.~Abazov {\it et al.} [D0 Collaboration], Nucl. Instrum. Methods
Phys. Res A {\bf 552}, 372 (2005).
\bibitem{Abazov:2008yf}
  V.~M.~Abazov {\it et al.}  [D0 Collaboration],
  arXiv:0808.0269 [hep-ex].
\bibitem{GMSBref}
V.~M.~Abazov {\it et al.}  [D0 Collaboration],
  arXiv:0710.3946 [hep-ex].

\bibitem{eelpaper}
D\O\ Collaboration, 
D\O-Note 5464-CONF (Aug. 2007), \\http://www-d0.fnal.gov/Run2Physics/WWW/results/prelim/NP/N57/N57.pdf;  D\O\ Collaboration,D\O-Note 5127-CONF (June 2006), http://www-d0.fnal.gov/Run2Physics/WWW/results/prelim/NP/N46/N46.pdf.
\bibitem{mumulpaper}
D\O\ Collaboration, 
D\O-Note 5348-CONF (March 2007),\\ http://www-d0.fnal.gov/Run2Physics/WWW/results/prelim/NP/N52/N52.pdf.
\bibitem{mumupaper}
D\O\ Collaboration, 
D\O-Note 5126-CONF (June 2006),\\ http://www-d0.fnal.gov/Run2Physics/WWW/results/prelim/NP/N45/N45.pdf.

\end{thebibliography}
\end{document}